\documentclass[times,twocolumn,final]{elsarticle}

\usepackage{amsmath,amssymb,amsfonts,bm}
\usepackage{placeins}
\usepackage{relsize}
\usepackage{float}
\usepackage{pifont}
\usepackage{balance}
\usepackage[numbers]{natbib}
\usepackage{algorithm}
\usepackage{algpseudocode}
\usepackage{tabularx,booktabs}
\usepackage[inline]{enumitem}
\usepackage[colorlinks=false]{hyperref}

\newcommand{\slicemodule}{\textit{Slice Module }}
\newcommand{\srmodule}{\textit{SR Module }}
\newcommand{\cmark}{\ding{51}}%
\newcommand{\xmark}{\ding{55}}%
\newcolumntype{C}{>{\centering\arraybackslash}X} 

\usepackage{framed,multirow}

\usepackage{amssymb}
\usepackage{latexsym}
\usepackage{url}
\usepackage{xcolor}

\definecolor{newcolor}{rgb}{.8,.349,.1}

\begin{document}

\begin{frontmatter}

\title{Meta-learning Slice-to-Volume
Reconstruction in Fetal Brain MRI using Implicit Neural
Representations}%

\author[1]{Maik Dannecker \corref{cor1}}
\cortext[cor1]{Corresponding author:}
\ead{m.dannecker@tum.de}
\author[2]{Thomas Sanchez}
\author[2]{Meritxell Bach Cuadra}
\author[1]{Özgün Turgut}
\author[3]{Anthony N. Price}
\author[4]{Lucilio Cordero-Grande}
\author[3]{Vanessa Kyriakopoulou}
\author[3]{Joseph V. Hajnal}
\author[1,5]{Daniel Rueckert}

\address[1]{School of Computation, Information and Technology, and School of Medicine and Health, Technical University Munich, Munich, Germany}
\address[2]{CIBM Center for Biomedical Imaging and Radiology Department of Lausanne University Hospital and University of Lausanne, Lausanne, Switzerland}
\address[3]{Centre for the Developing Brain, School of Biomedical Engineering and Imaging Sciences, King’s College London, London, United Kingdom}
\address[4]{Biomedical Image Technologies, ETSI Telecomunicación, Universidad Politécnica de Madrid, Madrid, Spain}
\address[5]{Department of Computing, Imperial College London, London, United Kingdom}


\begin{abstract}
High-resolution slice-to-volume reconstruction (SVR) from multiple motion-corrupted low-resolution 2D slices constitutes a critical step in image-based diagnostics of moving subjects, such as fetal brain Magnetic Resonance Imaging (MRI). Existing solutions struggle with image artifacts and severe subject motion or require slice pre-alignment to achieve satisfying reconstruction performance. We propose a novel SVR method to enable fast and accurate MRI reconstruction even in cases of severe image and motion corruption. Our approach performs motion correction, outlier handling, and super-resolution reconstruction with all operations being entirely based on implicit neural representations. The model can be initialized with task-specific priors through fully self-supervised meta-learning on either simulated or real-world data. In extensive experiments including over 480 reconstructions of simulated and clinical MRI brain data from different centers, we prove the utility of our method in cases of severe subject motion and image artifacts. Our results demonstrate improvements in reconstruction quality, especially in the presence of severe motion, compared to state-of-the-art methods, and up to 50\% reduction in reconstruction time.
\end{abstract}


\end{frontmatter}


\section{Introduction}
\label{sec:intro}
\subsection{Motivation}
\label{subsec:motiv}
3D MRI scans of the human brain provide valuable insights for medical diagnosis and research. To achieve high-quality corruption-free imaging, any motion during the acquisition process should be kept to a minimum. This is especially challenging for MRI in fetal subjects, where minimization of subject motion is impossible. Acquisition sequences like single-shot fast spin echo (SSFSE) have been developed to acquire 2D slices in under a second, virtually freezing motion in time \citep{prayer2004fetal, saleem2014fetal}. While this diminishes motion corruption within a slice, it does not address the problem of inter-slice motion \citep{gholipour2014fetal}. Moreover, the acquisitions come with slice thicknesses between 2 to 5 mm to obtain a reasonably high signal-to-noise ratio \citep{uus2023retrospective}. To gain sufficient information for the reconstruction process, multiple orthogonal stacks of slices are typically acquired \citep{uus2023retrospective, price2019dhcpACQ}. Reconstructing a high-resolution 3D brain from these motion-corrupted stacks of thick slices requires accurate slice re-alignment combined with super-resolution (SR). This process, termed slice-to-volume reconstruction (SVR), is usually approached in an iterative process of inter-slice motion correction interleaved with SR steps \citep{Rousseau2006, Gholipour2010, Jiang2007, Kuklisova2012, tourbier2015efficient, ebner2020automated}. The high-resolution reconstruction volume is thereby represented as a discrete 3D grid which comes with the drawback of high computational complexity and long run times. Recent approaches leverage implicit neural representations (INRs) \citep{Mildenhall2021, Sitzmann2020} for SR reconstruction \citep{Wu2021, Xu2023} and have shown to be computationally effective. However, these approaches come along with deficits in motion correction. In the case of \citet{Wu2021}, the assumption is limited to motion occurring between stacks, neglecting motion between slices. On the other hand, \citet{Xu2023} consider inter-slice motion but require an external registration step \citep{svort} trained in a supervised manner to pre-align slices with severe motion corruption to achieve satisfying reconstruction quality.
\subsection{Contribution}
\label{subsec:contrib}
We propose a minimal SVR setup based on the implicit neural representation (INR) of two multilayer perceptrons (MLPs) with sine non-linearities, known as SIREN \citep{Sitzmann2020}. Different to recently proposed methods \citep{Xu2023, Wu2021} that apply INRs to model super-resolution with INRs, we introduce the first fully INR-based SVR model, simultaneously addressing motion correction, outlier handling, and super-resolution of 2D slices into a 3D volume. Our key contributions can be summarized as follows:

\begin{itemize}
    \item We propose a fully INR-based SVR setup with self-supervised meta-learning, achieving faster and more robust slice-to-volume reconstruction — reducing computation times by up to 50\% compared to the fastest baselines \citep{Xu2023, Kuklisova2012} — while enhancing reconstruction performance under challenging conditions like extreme motion corruption or image artifacts.
    \item Contrary learning-based methods \citep{young2024fully, svort}, our meta-learning is fully \emph{self-supervised}, eliminating the need for synthetic data or minimally motion-corrupted acquisitions for high-quality ground truth generation \citep{svort}.
    \item We demonstrate the effectiveness of meta-learning under low data conditions with
    training sets of just 10 data samples, making it feasible for hospitals to train on in-house data circumventing potential domain shifts among datasets. 
    \item Finally, we validate our method on three datasets from two independent centers, evaluating over 480 reconstructions of more than 160 fetal subjects.
\end{itemize}
\section{Related Work}
\label{sec:related_work}
\subsection{Optimization-Based Methods}
An initial approach for SVR on fetal brains by \citet{Rousseau2006} used an iterative scheme of slice-to-volume registration followed by SR reconstruction. The authors employed normalized mutual information \citep{NMI_STUDHOLME199971} as a similarity metric to guide the slice registration. For SR, they approximated the point spread function (PSF) with Gaussian weighted scattered interpolation. \citet{Jiang2007} introduced cubic b-splines with multi-resolution control points to model the PSF. Oversampling on thin slices, this approach yielded sharper images, capturing fine anatomical details in the fetal brain. \citet{Gholipour2010} proposed a novel error norm using an M-Estimator demonstrating higher robustness to outliers, reducing the detriment of corrupted and misaligned slices. The concept of outlier rejection was adopted by\citet{Kuklisova2012} and further extended to a SR reconstruction method with complete outlier removal based on robust statistics. The complete rejection of corrupted slices furnished further improvements in reconstruction quality. Proposing Total Variation for SR regularization, \citet{tourbier2015efficient} attained increased robustness of their SVR method to motion artifacts. \citet{kainz_2015} accelerated SVR by utilizing multiple GPUs, with speed-up factors of up to 30 compared to single CPU systems. Finally, in 2020, \citet{ebner2020automated} presented a fully automatic framework for brain localization, segmentation, and reconstruction with outlier rejection of slices, based on the similarity to their simulated correspondents. Most of these tools, however, still struggle when confronted with severe subject motion \citep{uus2023retrospective}.

\subsection{Learning-Based Methods}
Learning based methods for SVR are trained neural networks that predict the slice transformations for motion correction either in a single forward pass \citep{young2024fully, yeung2021learning} or in an alternating approach of motion estimating and SR steps \citep{svort}. While convolutional neural networks are often the architecture of choice to predict the slice transformation parameters for motion correction \citep{hou20183, hou2018computing, salehi2018real, pei2020anatomy, yeung2021learning, young2024fully}, other approaches like SVoRT \citep{svort} employ transformer networks. SR steps are performed using classic interpolation methods or a trained interpolation network \citep{svort, young2024fully} is used to regress missing intensities of the 3D reconstruction. A limitation of learning-based methods is the need for supervised training data. Due to the inherent lack of ground truth data in the domain of slice-to-volume reconstruction, any supervised training requires a synthetically generated training set. High-quality 3D reconstructions of cases with little motion or brain atlases serve as ground truth from which 2D slices are extracted and rigidly transformed to simulate motion corruption \citep{Xu2023, svort}. The requirement of supervised synthetic training data introduces further limitations, including an artificially generated motion spectrum that might not adequately model fetal and maternal motion in the real world and the lack of robustness to domain shifts between training and real-world data, like different modalities, adult instead of fetal brains, different anatomy, or abnormal anatomy. These scenarios require further fine-tuning steps and additional training data to adequately fit the model to these new domains \citep{singh2020deep, zhang2024motion}. 

\subsection{Implicit Neural Representations}
\label{subsec:inrs}
Previous optimization-based methods and learning-based methods represent images as discretized signals on a 3D grid. However, input data is usually not acquired on a regular 3D grid \citep{Jiang2007, Xu2023}, complicating the interpolation for SR reconstruction. Moreover, for 3D reconstruction, computational complexity increases cubically with higher resolution, which can lead to reconstruction times exceeding two hours for optimization based methods \citep{uus2023retrospective}. Implicit neural representations (INRs), on the other hand, model the data as a continuous function of spatial coordinates, solving the problem of irregularly sampled data and facilitating a resolution agnostic reconstruction.
INRs have gained popularity when \citet{Mildenhall2021} proposed a method to synthesize 3D scenes by optimizing an underlying continuous function modeled by a multilayer perceptron (MLP). The MLP is fitted to acquired views of the scene by feeding it a 5-dimensional input (spatial location and viewing direction) and predicting volume density and radiance. After the optimization process, the model can interpolate realistic unseen views of the scene. \citet{tancik2020fourfeat} extended the work by introducing positional Fourier-Encodings, which map the coordinate inputs to a higher-dimensional space. This improves convergence speed and allows the network to better capture fine image details, i.e., high-frequency signals. Other approaches employ learnable hash encodings to achieve fast reconstructions while maintaining high quality \citep{mueller2022instant}. Instead of using any encodings, \citet{Sitzmann2020} introduced sinusoidal representation networks, dubbed SIREN, replacing the ReLU activation functions of the MLP by a sine activation. Combined with a specific initialization scheme, they achieved fast convergence and high-quality reconstructions.
\subsubsection{INRs for Medical Image Reconstruction}
\citet{Wu2021} presented an adaptations of INRs for medical image reconstruction, achieving high-resolution volume reconstruction from low-resolution slices of MRI acquisition stacks. The authors employ an initial stack alignment but assume the absence of any inter-slice motion. For SR reconstruction, they use an MLP with positional Fourier-Encoding~\citep{tancik2020fourfeat} to learn a continuous high-resolution volumetric representation of the brain. Introducing Neural Slice-to-Volume Reconstruction (NeSVoR), \citet{Xu2023} specifically aim for INR-based volume reconstruction from low-resolution stacks with inter-slice motion corruption. The authors employ a hash-grid encoded MLP \citep{mueller2022instant} for SR, bias-field estimation, and noise regularization. During reconstruction, the model simultaneously learns the SR reconstruction and slice re-alignment, represented as a 6-dimensional learnable parameter per slice. Whereas NeSVoR can adequately handle small to moderate inter-slice motion, it fails for cases with severe motion corruption, relying on initial slice pre-alignment from learning-based methods like SVoRT \citep{svort}. This re-introduces the aforementioned drawbacks of learning-based methods, e.g., the need of supervised synthetic training data and the susceptibility to domain shifts.

\subsection{Meta-Learning}
In a typical INR based image reconstruction setup, the representation network is (re-)initialized and trained from scratch for any new input signal, without leveraging any domain-specific prior knowledge. Various approaches to incorporate such priors into INRs exist, including the concatenation of a latent vector to the input coordinates \citep{Park_2019_CVPR}, or the application of a hypernetwork that maps an input signal to INR weights \citep{sitzmann2019srns}. Meta-learning \citep{MetaLearningReview2022} offers another promising direction. In the context of few-shot learning, meta-learning helps the model to quickly adapt to unseen tasks with only a few training samples. For INRs, on the other hand, meta-learning yields faster convergence times and better reconstruction performance \citep{tancik2020meta, Yuce_2022_CVPR}. For example, MetaSDF \citep{NEURIPS2020_MetaSDF} employs an INR to learn shape representations as signed distance functions (SDFs) and achieves enhanced accuracy through meta-learning. This is particularly relevant for partial observations, such as sparse input signals, which commonly occur during SR of motion-corrupted slices in SVR.

Meta-learning for INRs comprises an inner- and outer optimization loop \citep{finn2017MAML, nichol2018first}. First, the (standard initialized) INR is duplicated to get an outer- and multiple inner models. Next, the inner models are optimized for one task each. The outer model (also called meta model) is then updated with a gradient step in the direction of the weights of the optimized inner models. Thereby, the outer model learns a weight initialization to efficiently optimize for new tasks of the underlying domain. While Model-Agnostic Meta-Learning (MAML) \citep{finn2017MAML} uses both first- and second-order gradients for updating the outer model, approaches like Reptile \citep{nichol2018first} rely on first-order gradients only. In the case of INRs, where a model is trained on large input signals (e.g., 3D MR images), Reptile is more suitable as computationally heavy second-order gradients are not required. This allows for more iterations in the inner loop, which facilitates more accurate modeling of the input signal by the inner model~\citep{tancik2020meta}.

\section{Methodology}
\begin{figure*}
\centerline{\includegraphics[width=\textwidth]{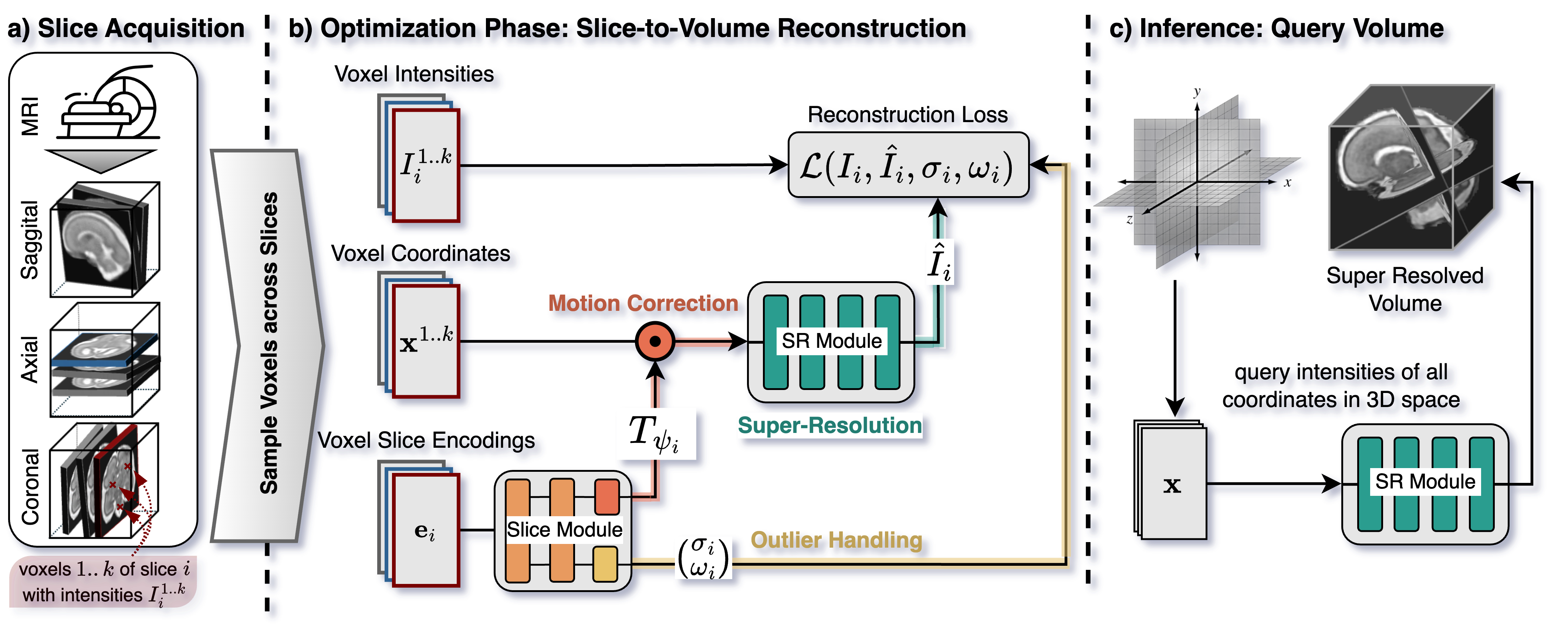}}
\caption{\unboldmath{Overview 3D reconstruction. a) Acquisition stacks of motion corrupted 2D slices across different anatomical planes. 
b) Architecture: Our framework comprises two key components: the \slicemodule and the \srmodule. The \slicemodule operates on positional encodings of slices, termed slice encodings, and is designed with two output heads. One head estimates motion correction while the other head addresses outlier handling, including slice intensity scaling ($\sigma$) and slice weighting ($\omega$). Simultaneously, the \srmodule learns a continuous super-resolved and \emph{motion-free} 3D brain representation by predicting voxel intensities of coordinates sampled from the motion corrected slices. 
c) After optimization, we obtain the high-resolution 3D reconstruction by querying the \srmodule with coordinates sampled from a 3D grid.}}
\label{main_architecture}
\end{figure*}
\label{sec:method}
We aim to reconstruct an unknown high-resolution volume \(V\) from multiple stacks of low-resolution 2D slices. To avoid notational burden, we will treat the different stacks as a single stack of 2D slices, where \(X_i\) denotes the \(i^{th}\) 2D slice. \(T_i\) accounts for the motion of slice $i$, and \(B_i\) represents the point spread function (PSF) of the acquisition sequence. These parameters connect the unknown volume \(V\) to each acquired slice \(X_i\) through a forward model \(X_i = B_iT_iV+\epsilon_i\), with \(\epsilon_i\) representing residual noise from the acquisition process. The unknown volume can then be estimated by solving an ill-posed inverse problem. While the PSF is assumed to be an anisotropic 3D Gaussian \citep{Rousseau2006, Jiang2007}, the slice motion \(T_i\) is unknown and needs to be estimated along with the high-resolution volume \(V\). 

\subsection{Proposed Model}
\label{subsec:model}
We propose a minimal SVR setup, where two MLPs, namely a \slicemodule and a \srmodule assume the task of motion correction with outlier handling and SR reconstruction, respectively. Both MLPs are designed with sine activation functions instead of ReLUs, dubbed sinusoidal representation networks (SIREN) in literature \citep{Sitzmann2020}. Contrary to the recently proposed SVR model NeSVoR \citep{Xu2023} where motion correction is modeled as learnable slice-parameters, we dedicate a single network, i.e., \slicemodule, to handle all slice specific operations, including outlier handling and motion correction. This design leads to a compact architecture of just two SIRENs using less than 20\% of NeSVoR's parameters, and improves reconstruction robustness, especially to motion-corruption. Additionally, as both SR and motion correction are performed by SIRENs, they equally benefit from the meta-learned initialization, introduced in Section \ref{subsec:meta_learning}. An overview of our SIREN based SVR design, dubbed (SSVR), is presented in Fig.~\ref{main_architecture}. 

The unknown high-resolution reconstruction volume $V:~\mathbb{R}^4~\rightarrow \mathbb{R}$ is represented by the \srmodule which operates on the homogeneous coordinates \(x=(x_1,x_2,x_3,1)\). Each slice $i$ is assigned a positional encoding, here termed slice encoding, $e_i=(stack_{idx}, slice_{idx})$. Given the slice encoding $e_i$, the \slicemodule estimates the motion correction modeled as rigid transformation and parametrized as $T_{\psi_i} \in \mathbb{R}^{4\times4}$ with six degrees of freedom: three rotations and three translations. Additionally, via a second output head, the \slicemodule performs outlier handling parameterized by $\sigma_i$ and $\omega_i$ for slice intensity scaling and slice weighting (applied in Eq. \ref{eq:outlier_handling}), respectively.  
The PSF is modeled as an anisotropic 3D normal distribution \citep{Rousseau2006, Kuklisova2012} with a covariance matrix $\Sigma_i$ defined as in \citep{Jiang2007}. We can then formulate a continuous forward model 
\begin{equation}\label{eq:forward continuous}
{I}_i(x) = \sigma_i \int_{\mathcal{N}(x, \Sigma_i)} V\left(T_{\psi_i} x\right) + \epsilon_i\left(T_{\psi_i} x\right) \, dx,
\end{equation}
where $\sigma_i$ is the intensity scaling for slice $i$ that accounts for global inhomogeneities across slices, and $\epsilon_i\left(T_{\psi_i}(x)\right)$ is the slice dependent zero-mean Gaussian noise \citep{Xu2023}. $I_i(x)$ is a random variable denoting the intensity of a pixel at homogeneous coordinate $x \in \mathbb{R}^4$ in slice $i$. Our setup does not include bias field correction, which is accounted for during post-processing \citep{N4_Tustison}. As the continuous forward model of Eq. \eqref{eq:forward continuous} has no closed-form solution, we need to approximate it in order to set up the objective that we want to minimize. We first compute the average pixel intensity as 
\begin{equation}\label{eq:SVR_continuous}
\mathbb{E}[I_i(x)] = \sigma_i \int_{\mathcal{N}(x, \Sigma_i)} V\left(T_{\psi_i} x\right) \, dx,
\end{equation}
and numerically approximate the integral using Monte Carlo sampling. We follow the approach of \citep{Xu2023} by sampling $K$ i.i.d. points, \(u_{ik} \sim \mathcal{N}(0, \Sigma_i)\) for \(k\in [1, K]\), and approximate Eq.\eqref{eq:SVR_continuous} by
\begin{equation}\label{eq:SVR_discrete}
\mathbb{E}[I_i(x)] \approx \widehat{I}_i(x) = \frac{\sigma_i}{K}\sum_{k=1}^{K} V\left(T_{\psi_i} (x+u_{ik})\right),
\end{equation}
where $\widehat{I}_i(x)$ can be viewed as a slice simulated from the volume \(V\). This yields an objective function that is minimized for $V$ and $T_{\psi_i}$ via gradient descent
\begin{equation}\label{eq:loss_function}
\mathcal{L} = \mathlarger{\sum_{i=1}^{N}} \frac{1}{\lvert X_i \rvert}\sum_{x\in X_i} \lvert I_i(x)-\widehat{I}_i(x) \rvert,
\end{equation}
denoting the mean absolute error between the intensities at coordinates $x \in X_{1...N}$ of the acquired slices $I_{1...N}$ and simulated slices $\widehat{I}_{1...N}$. 
Using the mean absolute error helps us to reduce the impact of outliers in the form of image corruptions, attaining a more robust reconstruction compared to the mean-squared error.

\subsection{\srmodule}
\label{subsubsec:sr_siren}
Representing the unknown high-resolution volume $V$, the \srmodule receives input in the form of motion corrected, homogeneous coordinates, $T_{\psi_i}(x) \in \mathbb{R}^4$, and predicts the corresponding intensities $\widehat{I}_i(x)$. The \srmodule thereby represents a continuous function of the form $\mathbb{R}^4 \rightarrow \mathbb{R}$. Unconstrained to a discrete grid, our model can innately handle the irregularly acquired image data from different acquisition slices and stacks. We optimize the \srmodule jointly with the \slicemodule via gradient descent by minimizing Eq. \eqref{eq:loss_function}. During inference, we can then generate the super-resolved reconstructed volume $V$ at a desired resolution by querying the \srmodule with the 3D coordinates of the respective volume (see Fig. \ref{main_architecture}c).

\subsection{Slice Module} 
\label{subsubsec:motion_correction}
The \slicemodule with its two output heads is responsible for all slice-specific operations, including motion correction and outlier handling. 
\subsubsection{Motion Correction} 
Given a unique slice encoding $e_i=(stack_{idx}, slice_{idx})$, the \slicemodule estimates a rigid transformation $T_{\psi_i} \in \mathbb{R}^{4\times4}$, represented by $\psi_i \in \mathbb{R}^6$ (three Euler angles + 3D translation) which aligns the motion corrupted slice $i$ with the 3D volume $V$. During reconstruction, accurate motion correction is learned by minimizing the reconstruction error in Eq. \eqref{eq:loss_function} jointly with the \srmodule. Designing the \slicemodule as MLP with sine activation functions, i.e. as SIREN \citep{Sitzmann2020}, allows the network to learn fast, accurate, and robust motion correction. Indeed, it has been shown by \citet{Sitzmann2020} that SIREN can fit a signal faster and more accurately compared to an MLP without a periodic activation function. Moreover, the authors show that SIREN is better capable of reconstructing high-frequency signals, in our case expressed as abrupt and severe motion between neighboring slices \citep{Sitzmann2020, tancik2020fourfeat}. The experiments in Section \ref{subsec:results_sim_data} and the results of Table \ref{tab:ablation_table} support this reasoning, showing superior performance when using SIREN compared to an MLP with ReLU activation functions or modeling slice transformations $\psi$ as learnable parameters as in \citep{Xu2023}. 
\subsubsection{Outlier Handling}
The second task of the \slicemodule involves outlier handling (yellow output head in Fig.\ref{main_architecture} b)). Due to subject motion during the acquisition process, acquired slices are often corrupted by image artifacts such as blurring, ghosting, signal dropout, or other inhomogeneities like global intensity differences. To prevent the propagation of these artifacts into the reconstructed volume, the \slicemodule estimates adaptive slice intensity scalings $\sigma$ to correct for global intensity differences of individual slices, and slice weights $\omega$ to attenuate the impact of slices on the reconstruction process. High-quality slices are assigned a high impact, severely corrupted slices, i.e. outliers, are assigned a low impact. Eq. \eqref{eq:loss_function} is extended to:
\begin{equation}
\label{eq:outlier_handling}
\mathcal{L} = \mathlarger{\mathlarger{\sum_{i=1}^{N}}} \frac{\omega_i}{\lvert X_i \rvert}\mathlarger{\sum_{x\in X_i}} \lvert I_i(x)-\sigma_i\widehat{I}_i(x) \rvert,
\end{equation} 
while constraining $\omega$ and $\sigma$ to:
\begin{equation}
\frac{1}{N}\sum_{i=1}^{N} \omega_i=1,
\;
\frac{1}{N}\sum_{i=1}^{N} \sigma_i=1
\end{equation}
using softmax as final activation function. Contrary to \citep{Xu2023}, where slice weights are estimated by the slice variance, in our approach the model learns slice scalings and weights end-to-end by minimizing Eq. \eqref{eq:outlier_handling}.

\subsection{Meta-Learning}
\label{subsec:meta_learning}
\subsubsection{Theory of Meta-Learning}
\label{subsec:meta_learning_theory}
Meta-learning aims at learning how to learn \citep{MetaLearningReview2022}. In the context of few-shot learning, meta-learning enables the model to adapt to unseen tasks quickly and with only a few samples. In the most prominent adaptations of meta-learning, the training process is built with an inner and outer optimization loop~\citep{finn2017MAML,nichol2018first}. 

In the inner loop, we optimize a model \(f\) with initial weights \(\theta^0_{\rho} = \theta\) and inner learning rate $\alpha$ for a task \(\rho \in \mathcal{P}\) by minimizing loss \(\mathcal{L}_{\rho}\) through \(n>0\) gradient steps,
\begin{equation}\theta^{k+1}_{\rho} = \theta^k_{\rho} - \alpha \nabla_{\theta^k_{\rho}} \mathcal{L}_{\rho}(f_{\theta^k_{\rho}}), \quad 0\leq k < n\label{metaeq1}\end{equation}
The objective of the meta-optimization is then to find an initialization \(\theta\) that minimizes the error on tasks \(\mathcal{P}\):
\begin{equation}\min_{\theta} \sum_{\rho \in \mathcal{P}}\mathcal{L}_{\rho}(f_{\theta^n_{\rho}})\label{metaeq2}\end{equation}
resulting in the meta-gradient step with outer learning rate $\beta$:
\begin{equation}\theta = \theta - \beta \, \nabla_{\theta} \sum_{\rho \in \mathcal{P}}\mathcal{L}_{\rho}(f_{\theta^n_{\rho}})\label{metaeq3}\end{equation}
This involves a second-order gradient, which makes it memory intensive, limiting the number of inner iterations.
Reptile \citep{nichol2018first} only computes the first-order gradient, making it less memory intensive and allowing for more inner iterations to better fit a task. This meta-learning technique is described by Algorithm~\ref{alg:meta_learning}.

\subsubsection{Meta-Learning Slice-to-Volume Reconstruction}
\label{subsec:meta_learning_inr_svr}
Meta-learning has already been successfully applied to INRs, where it is reported to speed up convergence and to improve reconstruction quality \citep{NEURIPS2020_MetaSDF, tancik2020meta, Yuce_2022_CVPR}. As our model is designed so that all reconstructions tasks, namely SR, motion correction, and outlier handling, are INR-based, we can fully capitalize on these findings. 

For MRI brain reconstruction, we employ meta-learning by training on a set of reconstruction tasks \(\mathcal{P}\). We refer to a task \(\rho \in \mathcal{P}\) as the 3D reconstruction of a number of motion-corrupted slice acquisitions from an arbitrary number of acquisition stacks. \emph{Inner model} refers to the architecture outlined in Fig. \ref{main_architecture} b), including the \slicemodule and the \srmodule, and \emph{meta weights} refers to the set of weights used to initialize the model. Similar to \citep{tancik2020meta}, we observed better performance using first-order methods as they allow for more inner steps, enabling the inner model to better fit the underlying signal. By running Algorithm \ref{alg:meta_learning}, the meta-weights $\theta$ converge to an optimized state, providing an initialization to the inner model that enables rapid and accurate reconstruction of the training tasks \(\mathcal{P}\). We monitor this behavior on hold-out validation tasks and stop training once performance on these tasks has reached its optimum. \\
The learned meta-initialization equips the model with an implicit prior of the task domain \citep{tancik2020meta}, in our case of brain MRI. This prior knowledge speeds up reconstruction (see Fig. \ref{fig:TimeSeries}), and importantly, boosts the model's robustness to severe motion corruption and challenging reconstruction cases in general, as demonstrated in Section \ref{sec:results}. \\
Excluding the performance and robustness improvements from meta-learning, the extra training time for meta-learning is compensated after a sufficient number of reconstructions. For instance, with the dHCP dataset (Section \ref{sec:dHCP_real_data}), meta-learning took 4 hours and reduced the mean reconstruction time by 60\%, breaking even after 49 reconstructions. This calculation does not account for any quality improvements from the meta-learned prior.
\begin{algorithm}
\caption{Meta-learning for 3D brain reconstruction}\label{alg:meta_learning}
\begin{algorithmic}[1]
\Require $\mathcal{P}$: set of reconstruction tasks
\Require $\alpha, \beta$: inner and outer learning rate 
\State Randomly initialize meta weights $\theta$
\While{not done}
\State Sample $\rho \in \mathcal{P}$
\State Initialize inner model with meta weights $\theta_{\rho} \gets \theta$
\For{$iteration = 1, 2, .... $}
\State Update \(\theta_{\rho} \gets \theta_{\rho} - \alpha \nabla\mathcal{L}_{\theta_{\rho}}(\rho)\)
\EndFor
\State Update meta weights \(\theta \gets \theta-\beta(\theta - \theta_{\rho})\)
\EndWhile
\end{algorithmic}
\end{algorithm}

\section{Datasets}
\label{subsec:datasets}

\subsection{Simulated Fetal Data} \label{dHCP_synthetic_data}
Using the fetal dHCP data \citep{price2019dhcpACQ}, we generated five increasingly corrupted datasets, with dataset 1 exhibiting minimal corruption, and dataset 5 showing extreme image and motion corruption. Each dataset consists of 20 cases, where each case includes three orthogonal acquisition stacks of 2D simulated slices with in-plane resolution of 1.125 x 1.125 mm, slice thickness of 3.3 mm, and no slice gap. All five datasets amount to 100 cases of 20 unique subjects. We simulated the motion corruption by applying a random 3D rigid transformation to each slice, where we sampled the 3D rotation parameters uniformly and independently from \(\mathcal{U}(-6\mu, 6\mu)\) degrees and the 3D translation from \(\mathcal{U}(-4\mu, 4\mu)\) mm with $\mu \in [1, 5]$ for datasets 1 to 5. Additionally, we randomly corrupted slices with image artifacts including motion blurring and ghosting, and added Gaussian noise and random changes of contrast \citep{perez-garcia_torchio_2021}. 

\subsection{Fetal Data from dHCP}
\label{sec:dHCP_real_data}
For experiments on real-world data, we selected a total of 121 subjects from the fetal dHCP dataset \citep{price2019dhcpACQ} spanning an age range of 20 - 36 weeks gestational age. Each subject includes 6 uniquely oriented motion corrupted stacks, acquired on a 3T Philips Achieva system with the following parameters: T2-weighted SSFSE, TE=250ms, in-plane resolution of 1.1 x 1.1 $\text{mm}^2$, slice thickness of 2.2mm and negative slice gap of -1.1mm. The full acquisition protocol is described in \citep{price2019dhcpACQ}.

\subsection{Fetal Data from CHUV}
\label{sec:CHUV_real_data}
To test the performance consistency of the SVR methods across different acquisition parameters, we included 40 fetal subjects from a second clinical dataset from Lausanne University Hospital (CHUV). This retrospective study was part of an approved larger research protocol from local ethics committee (CER-VD 2021-00124). The studies were conducted in accordance with the local legislation and institutional requirements. Each subject contains at least 6 and up to 20 motion corrupted stacks, acquired on two different Siemens 1.5T scanners, with the following parameters: T2-weighted HASTE, TR=1200ms, TE = 90ms, in-plane resolution of 1.1 x 1.1$\text{mm}^2$, a slice thickness of 3mm and a slice gap of $0.3 \text{mm}$. 24 subjects were acquired on a Siemens Aera (Gestational age=$26\pm4$ weeks; $n_{stacks}=10.6\pm 5.0$) and 16 subjects were acquired on a Siemens MAGNETOM Sola (Gestational age=$30\pm 3.5$ weeks; $n_{stacks}=8.6\pm 3.0$). 

\begin{figure}
\centerline{\includegraphics[width=\columnwidth]{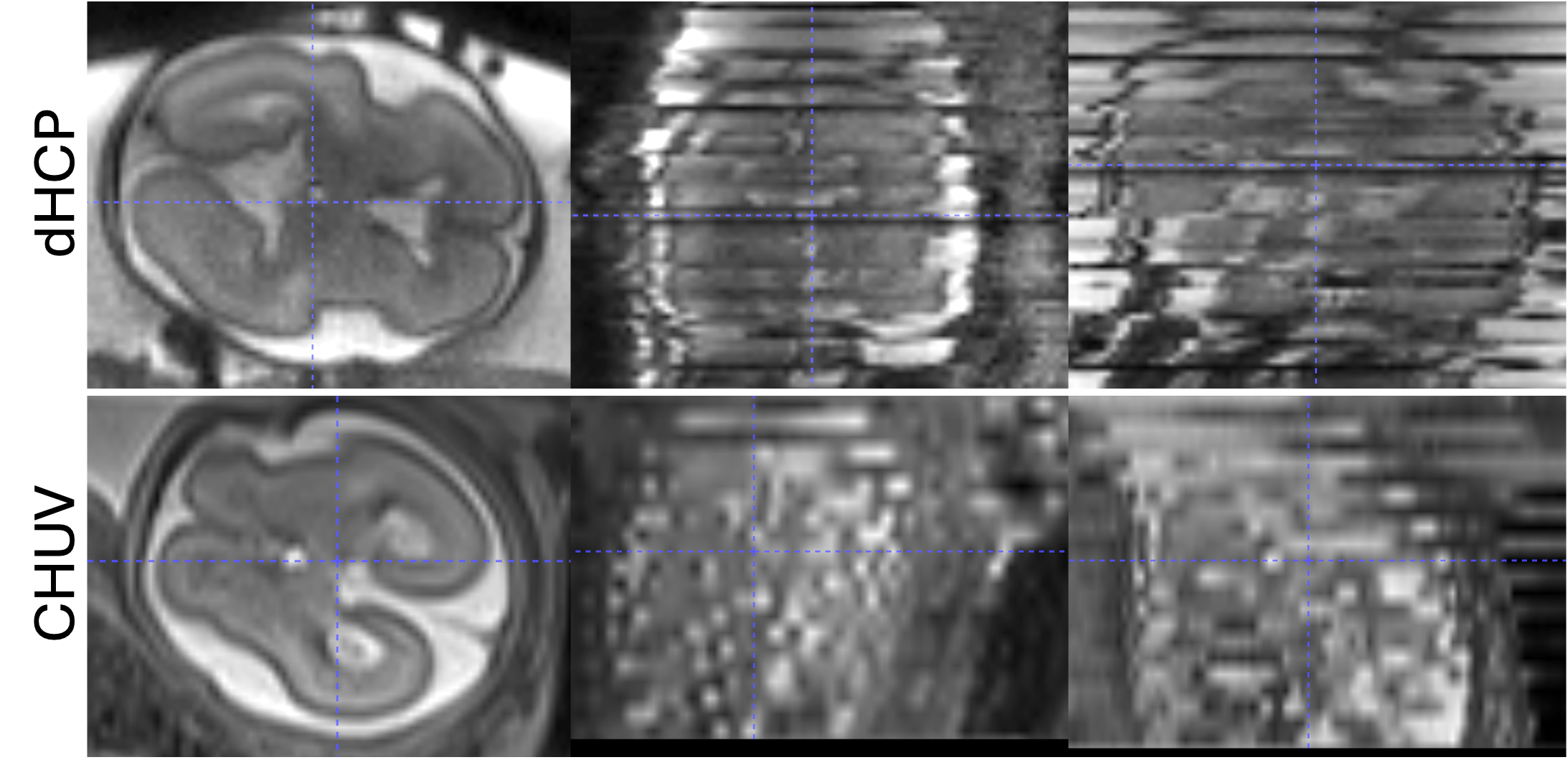}}
\caption{Real-world raw acquisition stacks from dHCP (top) and CHUV (bottom). Figure created with ITK-SNAP~\citep{itksnap}}  
\label{fig:acquisition_stacks}
\end{figure} 

\section{Experimental Setup}
\label{subsec:exp_setup}

\subsection{Pre- and Post Processing}
\label{subsec:preprocessing}
We applied automatic brain-masking \citep{ranzini2021monaifbs} for pre-processing, and normalized intensities to the interval $[0, 1]$. We applied bias field correction \citep{N4_Tustison} to the final reconstructions and reoriented them to the standard radiological space \citep{SVRTK_Reorient}. 
\subsection{Training Parameters}
\label{subsubsec:training_params}
We configured the \srmodule with six hidden layers, each containing 330 units, and the \slicemodule with two hidden layers of 256 units, and two output heads for motion correction and outlier handling. The learning rate $l_r$ for the \srmodule was set to $5\times 10^{-5}$, $l_r$ for \slicemodule was set to $5\times 10^{-4}$. For both models, we used CosineAnnealing as scheduler with a minimum learning rate of $2.5\times 10^{-5}$ and Adam as optimizer \citep{kingma2014adam}. The number of training iterations is adaptive to the total number of voxels of all acquisition slices and set according to the empirically determined equation to \(\textit{iteration}_{max} = \alpha\frac{\sum_{i=1}^{N} \lvert X_i \rvert}{\text{batchsize}}\). We set the batch size to 12000 and chose $\alpha=125$ when using a standard initialization and $\alpha=50$ when using a meta-learned initialization. The PSF was modeled as in \citep{Xu2023, Kuklisova2012, Jiang2007}, with covariance matrix $\Sigma=diag((\frac{1.2r_x}{2.355})^2,(\frac{1.2r_y}{2.355})^2,(\frac{r_z}{2.355})^2)$, given a slice resolution of $r_x\times r_y$ and slice thickness $r_z$. Different to \citep{Xu2023}, we set up $K$, i.e., the number of samples drawn to approximate the PSF, to quadratically increase with the number of iterations with an upper limit of 64, i.e., $K=max\left(1, 64 \times \left(\frac{\text{iteration}_{current}}{\text{iteration}_{max}}\right)^2\right)$. This speeds up the reconstruction time without sacrificing reconstruction quality. We ran all experiments on an NVIDIA RTX A6000 GPU. The setup amounts to a total of about $800\,000$ parameters.

\subsection{Meta-Learning Setup}
\label{subsubsec:meta_learning_setup} 
We conducted two meta-learning cycles to learn two meta-initializations. The first meta-learning was done on simulated data generated as described in Section \ref{dHCP_synthetic_data}. This dataset also consisted of ten subjects, each with six acquisition stacks corrupted by a random factor $\mu \in [1, 5]$. Again, 42 potential cases per subject were generated, and 20 cases were randomly selected, yielding 200 training cases. This meta-initialization was used for all experiments on simulated data and CHUV data, as both datasets have the same slice resolution of 1.125mm x 1.125mm x 3.3mm. This meta-learning run serves two purposes: First, it demonstrates that meta-learning can be conducted on data with simulated motion corruption. Second, using dHCP data with simulated motion corruption for training and testing the meta-learned initialization on clinical data from CHUV demonstrates the benefit of meta-learned initializations across domains. We refer to the model initialized with these meta-learned weights as $\text{SSVR}_{\text{meta1}}$.

The second meta-learning was done on dHCP data with a slice resolution of 1.1mm x 1.1mm x 2.2mm using ten subjects, each with six acquisition stacks. For each subject, we generated 42 potential cases by creating 20 permutations of three stacks, 15 permutations of four stacks, six permutations of five stacks, and one permutation of six stacks. We randomly sampled 20 permutations per subject, resulting in a training set of 200 cases. The final meta-initialization was used for all experiments on real-world dHCP data. We refer to the model initialized with these meta-learned weights as $\text{SSVR}_{\text{meta2}}$. 

Both meta-learning runs were validated on five subjects of the corresponding dataset using three acquisition stacks each. The validation process was self-supervised, based on the loss function in Eq. \ref{eq:loss_function}, requiring no ground truth. Each meta-learning training session took approximately four hours. None of the subjects used for evaluation in Section \ref{sec:results} were included in the meta-learning phase.

\subsection{Evaluation References} 
\label{subsec:evaluation_refs}
\textbf{Simulated Data}. For the evaluation on the dataset with simulated slices (Section \ref{dHCP_synthetic_data}), the 3D volume that was originally used to create the simulated acquisition stacks served as the reference for reconstruction, as it represents the ground truth.

\textbf{Real-world Data}. For the real-world datasets dHCP and CHUV, ground truth data is inherently unavailable. Therefore, we used a combination of slice pre-alignment followed by an optimization-based SVR method to generate the references. Specifically, we selected the learning-based pre-alignment tool SVoRT \citep{svort} and the optimization-based SVR method NeSVoR \citep{Xu2023}, a combination that has shown high robustness and proofed suitable for our experiments. All references and reconstructions have a $0.5$ mm isotropic resolution.

\subsection{Evaluation Metrics} 
First, we registered all reconstructions to the corresponding reference volume, described in Section \ref{subsec:evaluation_refs} to ensure best alignment. Next, we computed three image similarity metrics, namely peak signal-to-noise ratio (PSNR), structural similarity index measure (SSIM) \citep{SSIM}, and normalized cross correlation coefficient (NCC).

\subsection{Baselines}
\label{subsubsec:baselines}
We compared our method to two state-of-the-art, optimization based, open-source SVR methods: 
\begin {enumerate*}[label={(\arabic*)}]
    \item NeSVoR \citep{Xu2023}: Also based on INRs, NeSVoR is GPU-compatible and shows fast execution times. After hyper-parameter tuning on two subjects of each dataset, we found the default reconstruction parameters of NeSVoR to work best. However, to prevent under- and overfitting, we adapted the number of reconstruction iterations to the number of slices per stack and the number of acquisition stacks used as input. Thus, the number of iterations was set between 3000 (three acquisition stacks with few thick-slices) and 8000 (six acquisition stacks with many thin-slices). NeSVoR has approximately $4.8$ million parameters optimized during reconstruction, six times more than our proposed model.
    \item SVRTK: The optimization-based slice-to-volume reconstruction toolkit is the implementation of \citep{Kuklisova2012,uus2020deformable}. It offers multi-core processing, resulting in relatively fast reconstruction times. For our experiments we ran SVRTK with 32 CPUs in parallel. We encountered problems with cut-off regions in the final reconstruction when using SVRTK, which can be attributed to incomplete brain masks. This usually happens when the automated brain segmentation \citep{ranzini2021monaifbs} fails for slices with strong image corruption. Although we performed manual corrections for the SVRTK masks, occasional cut-offs in the final reconstructions remained.    
\end {enumerate*}

\section{Results} 
\label{sec:results}
\subsection{Results on Simulated Data}
\label{subsec:results_sim_data}
\begin{figure*}
\centerline{\includegraphics[width=\textwidth]{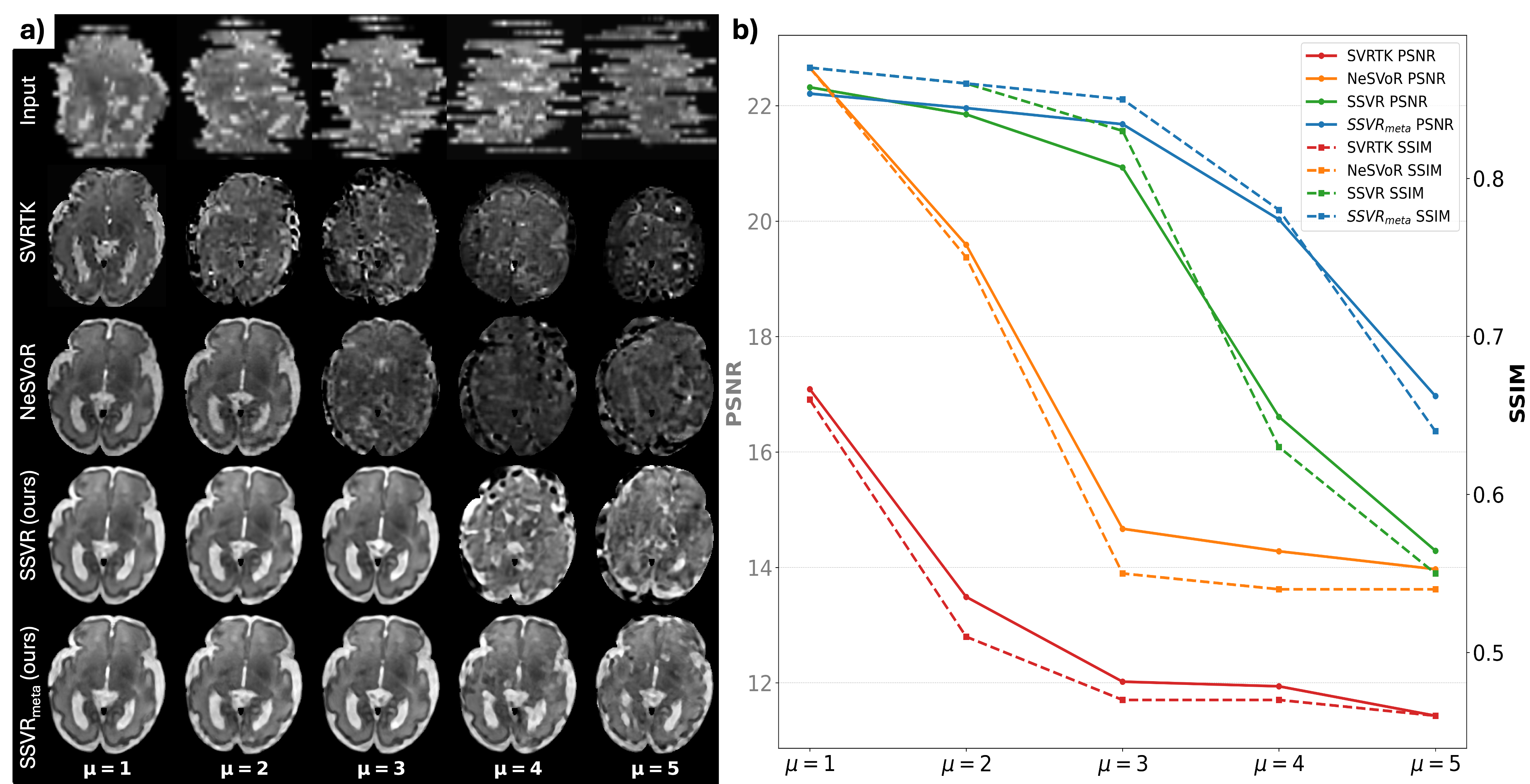}}
\caption{Evaluation of five simulated datasets with increasing simulated motion- and image corruption from left $(\mu=1)$ to right $(\mu=5)$. 2D visualization of 3D data. a) Reconstruction results of five cases with increasing simulated corruption. Top row shows one of three acquisition stacks used as input for the reconstruction process. Subsequent rows show reconstruction results of the baselines and our proposed method with standard initialization (SSVR) and meta-learned initialization ($\text{SSVR}_{\text{meta1}})$. b) Depicted are mean PSNR (solid lines) and SSIM (dashed lines) for the five simulated datasets with 20 cases each. X-axis represents the corruption factor applied to the dataset with $\mu=1$ indicating minimal corruption and $\mu=5$ extreme corruption (see Section \ref{dHCP_synthetic_data} for more details).}
\label{fig:sim_dhcp_corruption}
\end{figure*}

We quantitatively (Table \ref{table_results_sim_data}) and qualitatively (Fig. \ref{fig:sim_dhcp_corruption}) evaluated 3D brain reconstructions across five simulated datasets with increasing corruption of factors $\mu = 1$ (minimal corruption) up to $\mu = 5$ (extreme corruption). Our method performs similarly well as NeSVoR on minimally corrupted data of $\mu = 1$. However, with increasing corruption factors of $\mu=2$ and $\mu=3$, SVRTK and NeSVoR already show severe declines in reconstruction quality. Using a standard initialization, our method maintains stable performance and only shows performance drops when further increasing corruption to factors of $\mu=4$ and $\mu=5$. Using the meta-learned initialization mitigates these drops and yields the best reconstruction quality while reducing reconstruction time by 50\% compared to the standard initialization and by 66\% compared to NeSVoR.

\begin{table}
\caption{Evaluation of simulated data. Quantitative results of fetal brain reconstructions for 20 cases with minimally image- and motion corrupted acquisition stacks as input (top) and 20 cases with severely corrupted stacks (bottom). Mean values with standard deviation in parentheses. Best scores in \textbf{bold}, second best scores \underline{underlined}.}
\label{table_results_sim_data}
\small
\scriptsize
\begin{tabularx}{\columnwidth}{l c c c c}
\toprule
Method & PSNR $\uparrow$ & SSIM $\uparrow$ & NCC $\uparrow$ &Time (s) $\downarrow$\\
\midrule
\multicolumn{5}{c}{MINIMAL MOTION CORRUPTION $(\mu=1)$} \\
\midrule
SVRTK & \mbox{17.09 (1.45)} & 0.66 (0.06) & 0.87 (0.03) & 117 \\
NeSVoR & \mbox{\textbf{22.66 (1.36)}}  & \mbox{\textbf{0.87 (0.02)}}  & \mbox{\textbf{0.96 (0.01)}}  & 135 \\
SSVR (ours) & \mbox{\underline{22.32 (1.30)}} & \mbox{\textbf{0.87 (0.02)}} & \mbox{\textbf{0.96 (0.01)}} & \underline{96} \\
$\text{SSVR}_{\text{meta1}}$ (ours) & \mbox{22.21 (1.31)} & \mbox{\textbf{0.87 (0.02)}} & \mbox{\textbf{0.96 (0.01)}} & \textbf{45} \\
\midrule
\multicolumn{5}{c}{SEVERE MOTION CORRUPTION $(\mu=4)$} \\
\midrule
SVRTK & \mbox{11.94 (0.90)} & 0.47 (0.05) & 0.56 (0.05) & 168 \\
NeSVoR & \mbox{14.28 (0.77)} & 0.54 (0.04) & 0.75 (0.03) & 135 \\
SSVR (ours) & \mbox{\underline{16.61 (2.87)}} & \underline{0.63 (0.11)} & \underline{0.84 (0.07)} & \underline{96} \\
$\text{SSVR}_{\text{meta1}}$ (ours) & \mbox{\textbf{20.03 (1.39)}} & \mbox{\textbf{0.78 (0.05)}} & \mbox{\textbf{0.94 (0.05)}} & \textbf{45} \\
\bottomrule
\end{tabularx}
\end{table}

\subsection{Results on Fetal Data from the dHCP} 
\label{subsec:results_real_data}
\begin{figure}
\centerline{\includegraphics[width=\columnwidth]{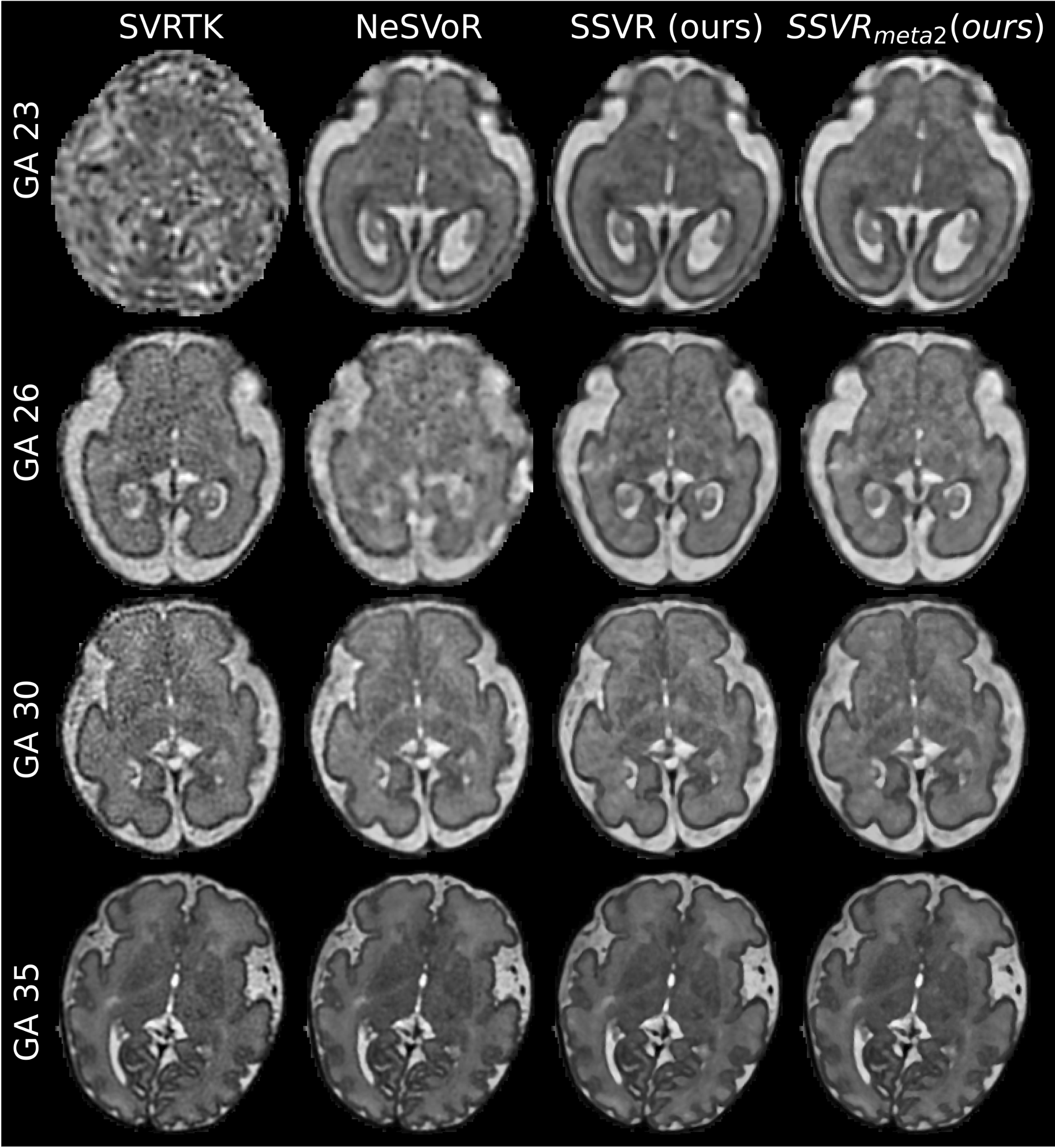}}
\caption{Reconstruction results of four cases of different weeks GA from the dHCP dataset. Columns show the results of each method. Baselines struggle particularly with younger cases which typically exhibit stronger motion corruption. Our method consistently produces high-quality reconstructions with standard (SSVR) and meta-learned initialization ($\text{SSVR}_{\text{meta2}}$).}
\label{fig:Exp1VisualShort}
\end{figure}

\begin{table*}
\caption{Evaluation of 40 fetal subjects of the dHCP data \citep{price2019dhcpACQ} (top) and of 40 fetal subjects from the CHUV data (bottom). Quantitative results for two setups. Left table: three acquisition stacks used as input for reconstruction, representing a challenging setup. Right table: all available acquisition stacks (6 for dHCP, $\geq 6$ for CHUV) used. Mean values with standard deviation in parentheses. Best scores in \textbf{bold}, second best scores \underline{underlined}.}
\label{tab:results_stack_ablation}
\small
\setlength{\tabcolsep}{4pt} 
\begin{tabularx}{\textwidth}{@{} l *{4}{C}  *{4}{C} @{}}
\toprule
& \multicolumn{4}{c}{3 INPUT STACKS} & \multicolumn{4}{c}{ALL ($\geq 6$) INPUT STACKS} \\
\cmidrule(lr){2-5} \cmidrule(lr){6-9}
Method & PSNR $\uparrow$ & SSIM $\uparrow$ & NCC $\uparrow$ & Time (s) $\downarrow$ & PSNR $\uparrow$ & SSIM $\uparrow$ & NCC $\uparrow$ & Time (s)$\downarrow$ \\
\midrule
& \multicolumn{8}{c}{dHCP Dataset} \\
\midrule
SVRTK & \mbox{18.68 (3.37)} & 0.68 (0.14) & 0.87 (0.10) & 281 & \mbox{22.01 (4.45)} & 0.79 (0.17) & 0.93 (0.09) & 563 \\
NeSVoR & \mbox{18.96 (2.37)}  & 0.70 (0.12)  & 0.90 (0.06)  & \underline{201} & \mbox{\textbf{26.78 (5.86})} & 0.87 (0.17) & \underline{0.96 (0.08)} & \underline{366}\\
SSVR (ours)   & \mbox{\underline{20.53 (2.67)}} & \underline{0.78 (0.10)} & \underline{0.93 (0.04)} & 240  & \mbox{26.51 (3.87)} & \mbox{\textbf{0.90 (0.11)}} & \mbox{\textbf{0.97 (0.07)}} & 486 \\
$\text{SSVR}_{\text{meta2}}$ (ours)  & \mbox{\textbf{21.51 (2.54)}} & \mbox{\textbf{0.81 (0.09)}} & \mbox{\textbf{0.94 (0.04)}} & \textbf{108} & \mbox{\underline{26.53 (4.11)}} & \mbox{\textbf{0.90 (0.11)}} & \underline{0.96 (0.15)} & \textbf{198} \\
\midrule
& \multicolumn{8}{c}{CHUV Dataset} \\
\midrule
SVRTK & \mbox{15.89 (2.46)} & 0.58 (0.14) & 0.78 (0.12) & \underline{118} & \mbox{16.68 (2.80)} & 0.65 (0.17)	& 0.81 (0.13) & \underline{246} \\
NeSVoR & \mbox{16.62 (2.97)} & 0.61 (0.15) & 0.80 (0.12) & 176 & \mbox{\textbf{22.73 (4.20)}} & \mbox{\textbf{0.85 (0.16)}} & 0.93 (0.16) & 295\\
SSVR (ours) & \mbox{\underline{19.72 (2.31)}} & \underline{0.75 (0.11)} & \underline{0.92 (0.6)} & 204  & \mbox{\underline{21.60 (2.06)}} & \mbox{\textbf{0.85 (0.08)}} & \mbox{\textbf{0.95 (0.04)}} & 363 \\
$\text{SSVR}_{\text{meta1}}$ (ours)  & \mbox{\textbf{19.90 (2.16)}} & \mbox{\textbf{0.77 (0.10)}} & \mbox{\textbf{0.93 (0.06)}} & \textbf{85} & \mbox{21.46 (2.29)} & 0.84 (0.10) & \mbox{\textbf{0.95 (0.05)}} & \textbf{222} \\
\bottomrule
\end{tabularx}
\end{table*}

We evaluated our method on 121 fetal subjects from the dHCP dataset spanning a gestational age of 21 to 38 weeks \citep{price2019dhcpACQ}. Each subject comes with six acquisition stacks of relatively thin slices with slice thickness of 2.2 mm and negative slice gap of 1.1 mm. The high amount of redundant information facilitates relatively easy reconstruction tasks. For an additional thorough analysis of the models' performance on more challenging tasks, we split the evaluation in two experiments: (i) Reconstruction and evaluation of 100 cases from 100 subjects using all six acquisition stacks as input. (ii) A \emph{stack-ablation} study on 40 subjects (19 from experiment 1 and 21 new) where the number of acquisition stacks used as input was gradually reduced to a minimum of three stacks, yielding 160 reconstruction tasks in total. The rationale behind experiment (ii) is that more input stacks provide greater redundancy, improving the SVR method’s ability to correct motion corruption and image artifacts. Conversely, fewer stacks increase the difficulty of the reconstruction task \citep{uus2023retrospective, Xu2023}. The evaluation references for both experiments were generated \emph{using all six acquisition stacks} as input and a combination of a slice pre-alignment tool \citep{svort} followed by the state-of-the-art optimization based SVR method NeSVoR \citep{Xu2023} for 3D reconstruction, as explained in Section \ref{subsec:evaluation_refs}. In summary, experiment (ii) allows for an objective measurement for robustness of the different methods despite the absence of an absolute ground-truth. A slower performance drop with a decreasing number of input stacks indicates a better robustness 

\textbf{Experiment (i): Reconstruction with All Stacks.}
Table \ref{tab:results_dhcp_100} shows the mean metrics for experiment (i). As expected, all methods provide similar reconstruction quality, with SVRTK performing the weakest. NeSVoR's high PSNR is due to the reference being generated by NeSVoR itself (after pre-alignment with SVoRT). Therefore, contrary to the other methods, NeSVoR encounters no domain shift between reconstruction and reference. Fig. \ref{fig:Exp1VisualShort} compares brain reconstructions of fetuses across different weeks GA. Here, baseline methods struggle particularly with younger cases. This could be due to stronger motion corruption as more space in the mother's womb allows for increased fetal movement. In contrast, our proposed method consistently produces high-quality reconstructions, even in the more challenging early-stage cases. Fig.~7 
of the supplementary material visualizes additional reconstructions of all discussed methods for qualitative evaluation.
\begin{table}
\caption[Ablation Study analyzing the effect of multimodal input, of spatial or $1D$ latent code $\bm{z}$, and of rigid subject alignment $\psi$ during training.]{\raggedright Ablation Study analyzing the effect of multimodal input, of spatial or $1D$ latent code $\bm{z}$, and of rigid subject alignment $\psi$ during training. We measured mean dice scores (DSC) and mean absolute error (MAE) of scan age (SA) and birth age (BA) in weeks. Best scores are marked bold.}

\caption{Evaluation of 100 fetal subjects of the dHCP data \citep{price2019dhcpACQ} using all six acquisition stacks for reconstruction. Mean values with standard deviation in parentheses. Best scores in \textbf{bold}, second best scores \underline{underlined}.}
\label{tab:results_dhcp_100}
\small
\scriptsize
\setlength{\tabcolsep}{4pt} 
\begin{tabularx}{\columnwidth}{l c c c c}
\toprule
Method & PSNR $\uparrow$ & SSIM $\uparrow$ & NCC $\uparrow$ &Time (s) $\downarrow$\\
\midrule
SVRTK & \mbox{23.81 (3.87)} & 0.88 (0.14) & 0.95 (0.08) & 525\\
NeSVoR & \mbox{\textbf{30.15 (5.97)}}  & \mbox{\textbf{0.93 (0.14)}}  & \mbox{0.96 (0.12)}  & \underline{366} \\
SSVR (ours)   & \mbox{27.90 (4.28)} & \mbox{0.92 (0.12)} & \mbox{\underline{0.97 (0.08)}} & 497 \\
$\text{SSVR}_{\text{meta2}}$ (ours) & \mbox{\underline{28.25 (3.43)}} & \mbox{\textbf{0.93 (0.09)}} & \mbox{\textbf{0.98 (0.05)}} & \textbf{202} \\
\bottomrule
\end{tabularx}
\end{table}

\begin{figure*}[ht]
\centerline{\includegraphics[width=\textwidth]{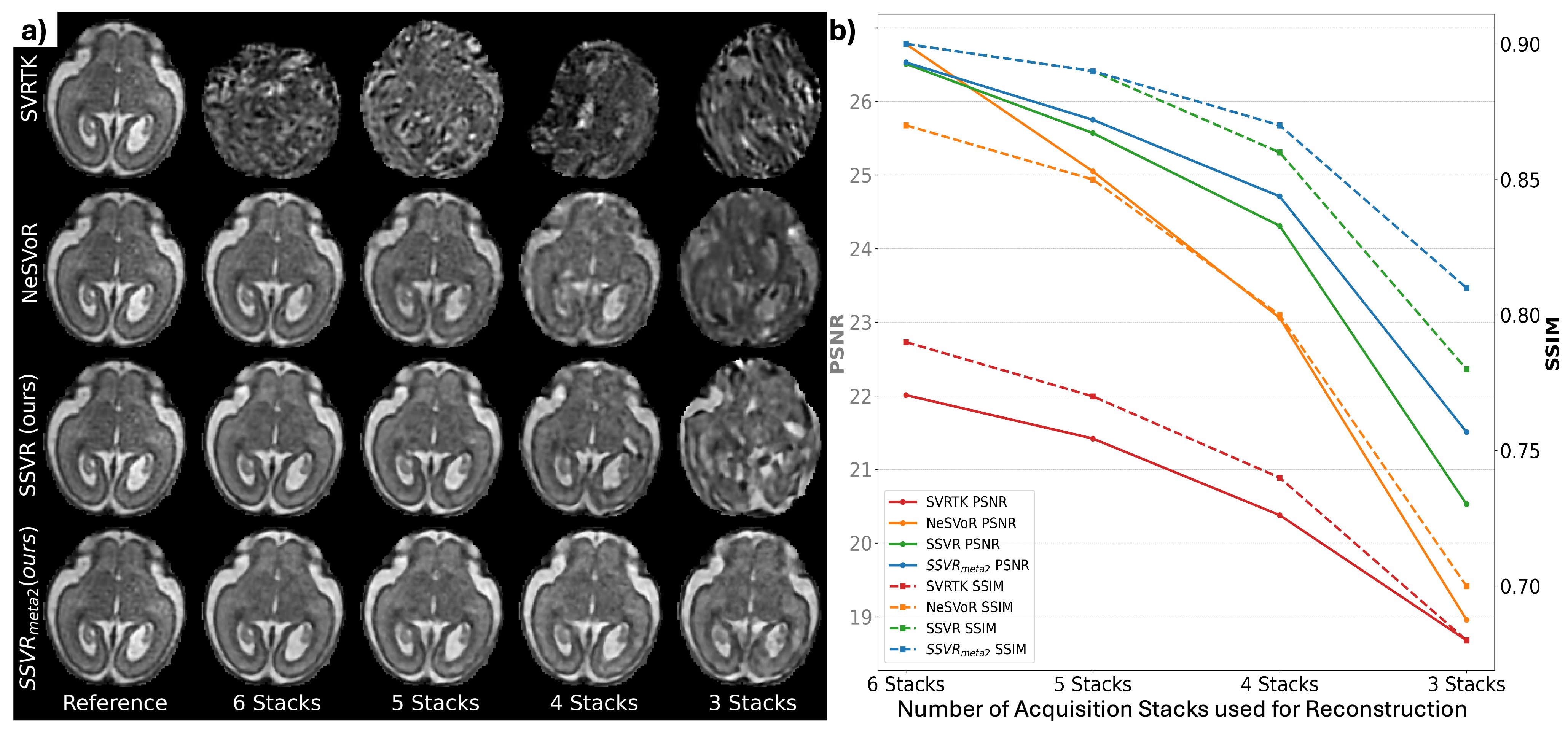}}
\caption{Stack ablation on 40 fetal subjects of the dHCP dataset. a) Leftmost column shows the reference generated as described in Section \ref{subsec:evaluation_refs}. Remaining columns present the reconstruction results for each method (rows) across the four stack ablation scenarios ranging from six, i.e., all available acquisition stacks, down to three stacks used for reconstruction. b) Mean PSNR (solid lines) and SSIM (dashed lines) of all methods for the different stack ablations counting 40 cases each.}
\label{dhcp_real_stack_ablation}
\vspace{-0.35cm}
\end{figure*}

\begin{figure}[ht]
\vspace{-0.1cm}
\centerline{\includegraphics[width=\columnwidth]{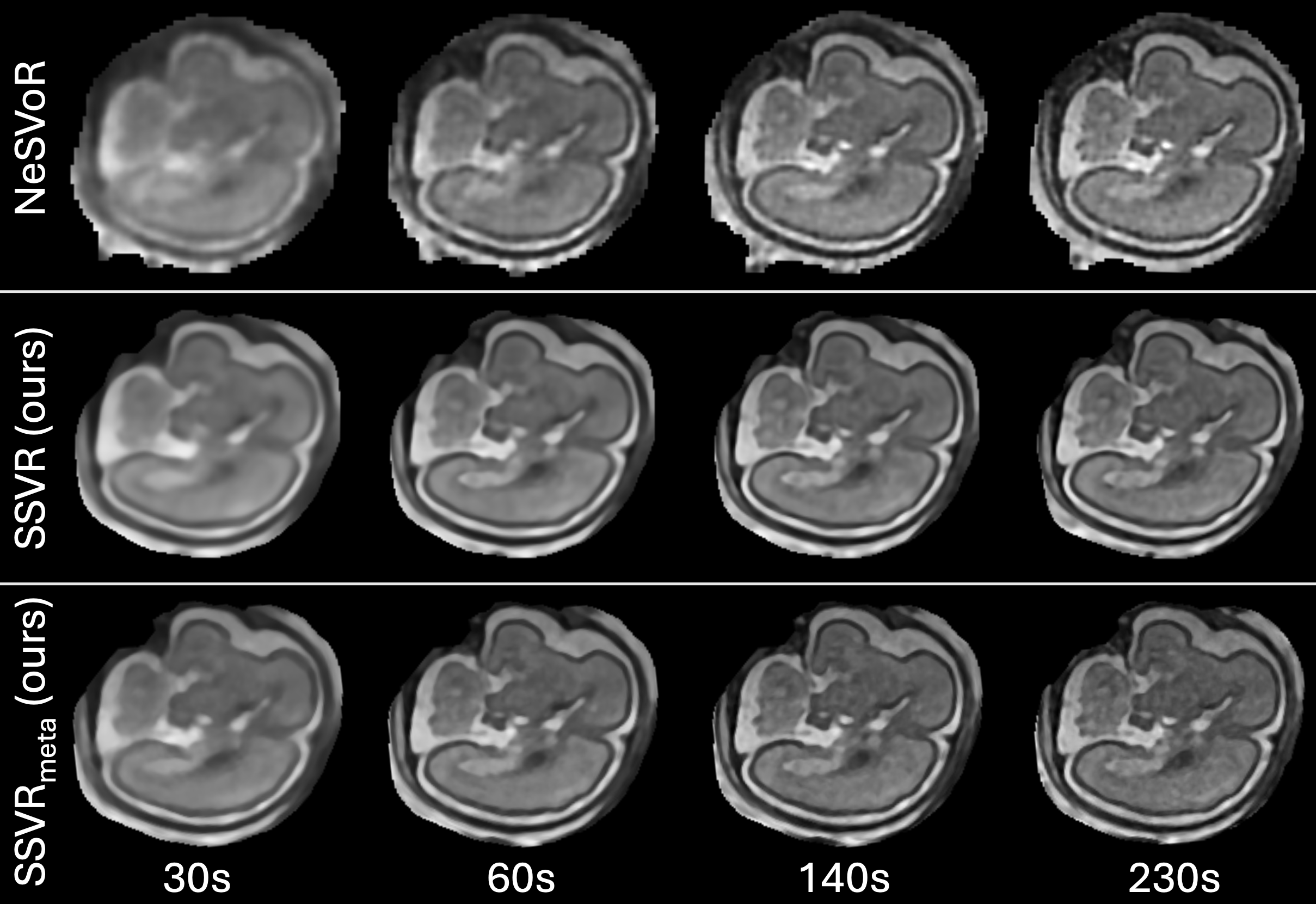}}
\caption{Reconstruction convergence. With standard initialization, our method converges quickly initially but takes longer to reconstruct fine details. Using a meta-learned initialization ($\text{SSVR}_{\text{meta2}}$), our method converges noticeably faster, surpassing NeSVoR.}  
\label{fig:TimeSeries}
\vspace{-0.70cm}
\end{figure}

\textbf{Experiment (ii): Stack Ablation.}
Table \ref{tab:results_stack_ablation} and Fig. \ref{dhcp_real_stack_ablation} provide an overview of the results of experiment (ii). The results agree with the findings for simulated data of Section \ref{subsec:results_sim_data} in that our method generally shows higher robustness and yields better reconstructions than the baselines when ablating stacks off the input. Again, a meta-learned initialization speeds-up the reconstruction time of our method (see Fig. \ref{fig:TimeSeries} and further boosts robustness compared to standard initializations. Fig. \ref{dhcp_real_stack_ablation} provides a qualitative example and mean PSNR and SSIM over all 40 cases for all stack ablations. SVRTK shows the weakest performance across all setups. Our proposed method and NeSVoR start similarly high. However, with decreasing number of input stacks, NeSVoR shows a faster and steeper drop in performance. With a standard initialization, our method achieves comparably good reconstruction quality. However, the meta-learned initialization again helps to further boost robustness, yielding reconstructions of highest quality among all setups. Remarkably, with just 3 acquisition stacks as input, we achieve a better mean SSIM score than SVRTK with all six stacks as input. In summary, our method demonstrates greater stability and smaller performance drops compared to the baselines. Using a meta-learned initialization, consistently yields the best results. 

\subsection{Results on Clinical Data from CHUV}
We have further evaluated the method on 40 cases of a clinical dataset from CHUV. Again, we have performed a stack ablation study, with the metrics presented in the lower half of Table \ref{tab:results_stack_ablation} and a qualitative example shown in Fig.~9 
of the supplementary material. The results corroborate the findings on the dHCP. Compared to the baseline, our method demonstrates the best reconstructions throughout the more challenging ablated setups (i.e. using just 3 or 4 stacks for reconstruction), and similar performance for the \emph{easier} setup, where all acquisition stacks are used as input (e.g., Fig.~8 
of the supplementary material). With meta-learned initialization, our method achieves the fastest reconstruction times requiring less than half the time of NeSVoR, and further improves reconstruction quality for the more challenging cases with just three acquisition stacks as input. 

\subsection{Ablation Study}
We have investigated the contribution of our proposals with an ablation study on 20 cases with strong ($\mu=3)$ simulated motion corruption. The study shows that using sine instead of ReLU as activation function of the \slicemodule (AF$_\text{SM}$) yields better motion correction and improves robustness to strongly corrupted data. Initializing the \slicemodule with a meta-learned initialization (SM$_\text{meta1}$) further strengthens the model's resilience to motion corruption. Outlier handling (OH) effectively handles image corruption and artifacts in slices, indicated by enhanced reconstruction quality. The full setup, employing meta-learned initializations for \slicemodule and \srmodule (SR$_\text{meta1}$) with outlier handling, yields the best performance. Finally, the meta-learned initialization reveals considerable speed-ups in reconstruction time (Fig. \ref{fig:TimeSeries}), reducing reconstruction times to approximately 40\% of that required with standard initialization, and about 50\% of the time needed by NeSVoR (see Table \ref{table_results_sim_data} and \ref{tab:results_dhcp_100}).

\begin{table}[H]
\caption{Mean metrics for ablation studies on 20 cases with strong simulated motion corruption ($\mu=3$). OH: Outlier Handling, $\text{AF}_\text{SM}$: Activation Function for \slicemodule, $\text{SM}_{meta}$: Meta-initialization for \slicemodule, $\text{SR}_{meta}$: Meta-initialization for \srmodule. Best scores in \textbf{bold}, second-best \underline{underlined}. Standard deviation in parentheses.}
\label{tab:ablation_table}
\small
\setlength{\tabcolsep}{6pt} 
\begin{tabularx}{\columnwidth}{l l l l c c}
\toprule
OH & $\text{AF}_\text{SM}$ & $\text{SM}_{meta}$ & $\text{SR}_{meta}$ & PSNR $\uparrow$ & SSIM $\uparrow$ \\
\midrule
\cmark & ReLU & \xmark & \xmark & 15.15 (0.66)  & 0.57 (0.04) \\
\cmark & Sine & \xmark & \xmark & 20.93 (2.16)  & 0.83 (0.08) \\
\xmark & Sine & \xmark & \xmark & 19.91 (2.86) & 0.78 (0.12) \\
\cmark & Sine & \cmark & \xmark & \underline{21.57 (1.30)}  & \underline{0.84 (0.03)} \\
\cmark & Sine & \cmark & \cmark & \mbox{\textbf{21.68 (1.29)}} & \mbox{\textbf{0.85 (0.03)}} \\
\bottomrule
\end{tabularx}
\end{table}

\section{Discussion and Conclusion}
We have introduced a novel slice-to-volume reconstruction (SVR) method using Implicit Neural Representations (INRs) for fetal brain MRI.Our work addresses remaining challenges of existing methods which often struggle with severe motion corruption and image artifacts. Different to alternative approaches, we advocate a fully INR-based architecture, utilizing sinusoidal representation networks, known as SIRENs, for motion correction, outlier handling, and super-resolution. This design demonstrates clear improvements in reconstruction quality and robustness compared to state-of-the-art baselines. On top of that, the design allows us to leverage recent insights on the benefits of meta-learning for INRs \citep{tancik2020meta, NEURIPS2020_MetaSDF, Yuce_2022_CVPR}. Incorporating optional \emph{fully self-supervised} meta-learning, we can further boost our model’s speed, robustness, and quality, cutting reconstruction times by 50\% while achieving remarkable resilience to motion corruption. Unsupervised meta-learning on just 10 real-world subjects already yields substantial performance gains, making it a practical solution in low data regimes, e.g., in clinical settings. \\

\textbf{Limitations and Future Work.} Despite the promising results, our method has some limitations which we will address in future work. The current outlier handling focuses on entire slices, which may be sub-optimal for slices with localized image corruption but which are otherwise of high quality. Future work will investigate pixel-wise outlier detection, which could offer more granular attenuation and improve reconstruction accuracy, as reported in previous work \citep{Kuklisova2012}. We recognize the importance of generalizability in clinical settings and will extend our meta-learning framework to adapt to the heterogeneity of acquisition sequences across different centers with a single meta-learned initialization. Lastly, brain analysis often requires downstream processes like tissue segmentation, a feature not currently integrated into existing SVR methods. Incorporating 3D segmentation reconstruction could provide complementary information, further guiding motion correction and enhancing the reconstruction quality \citep{stolt2023nisf}.

\section*{Funding and Acknowledgments}
This work was supported by the ERC (Deep4MI - 884622), and by the ERA-NET NEURON Cofund (MULTI-FACT - 8810003808), Swiss National Science Foundation grant 31NE30\_203977. We acknowledge the expertise of CIBM Center for Biomedical Imaging, a Swiss research center of excellence founded and supported by CHUV, UNIL, EPFL, UNIGE, and HUG.
\clearpage

\balance
\bibliographystyle{model2-names.bst}\biboptions{authoryear}
\bibliography{references.bib}

\end{document}